\documentclass[]{spie}  %>>> use for US letter paper
\usepackage{amsmath,amsfonts,amssymb}
\usepackage{graphicx}
\usepackage[colorlinks=true, allcolors=blue]{hyperref}

\title{Status of BICEP Array and Integration of the 220/270 GHz Receiver}

\author[a,*]{A.~Steiger}
\author[b]{P.~A.~R.~Ade}
\author[c,d]{Z.~Ahmed}
\author[e]{M.~Amiri}
\author[f]{D.~Barkats}
\author[a]{R.~Basu~Thakur}
\author[g]{C.~A.~Bischoff}
\author[h]{D.~Beck}
\author[a,i]{J.~J.~Bock}
\author[j]{V.~Buza}
\author[h,c]{B.~Cantrall}
\author[a]{J.~R.~Cheshire~IV}
\author[k]{J.~Connors}
\author[l]{J.~Cornelison}
\author[m]{M.~Crumrine}
\author[a]{A.~J.~Cukierman}
\author[k]{E.~Denison}
\author[n]{L.~Duband}
\author[f]{M.~A.~Echter}
\author[o]{M.~Eiben}
\author[f,p]{B.~D.~Elwood}
\author[a]{S.~Fatigoni}
\author[q]{J.~P.~Filippini}
\author[h]{A.~Fortes}
\author[a]{M.~Gao}
\author[g]{C.~Giannakopoulos}
\author[h]{N.~Goeckner-Wald}
\author[h]{D.~C.~Goldfinger}
\author[r,s]{S.~Gratton}
\author[h]{J.~A.~Grayson}
\author[a]{A.~Greathouse}
\author[f]{P.~K.~Grimes}
\author[e]{M.~Halpern}
\author[c,d]{S.~Henderson}
\author[m]{T.~D.~Hoang}
\author[k]{J.~Hubmayr}
\author[a]{H.~Hui}
\author[h]{K.~D.~Irwin}
\author[t]{M.~Izquierdo~Poza}
\author[a]{J.~H.~Kang}
\author[t]{K.~S.~Karkare}
\author[a]{S.~Kefeli}
\author[f,p]{J.~M.~Kovac}
\author[h]{C.~Kuo}
\author[m,u]{K.~Lasko}
\author[a]{K.~Lau}
\author[g]{M.~Lautzenhiser}
\author[h]{T.~Liu}
\author[j,v]{S.~C.~Mackey}
\author[m]{N.~Maher}
\author[i]{K.~G.~Megerian}
\author[a]{L.~Minutolo}
\author[a]{L.~Moncelsi}
\author[h]{Y.~Nakato}
\author[a,i]{H.~T.~Nguyen}
\author[a,i]{R.~O’Brient}
\author[f]{S.~N.~Paine}
\author[a]{A.~Patel}
\author[f]{M.~A.~Petroff}
\author[f,p]{A.~R.~Polish}
\author[n]{T.~Prouve}
\author[m]{C.~Pryke}
\author[k]{C.~D.~Reintsema}
\author[a]{T.~Romand}
\author[h]{M.~Salatino}
\author[a]{A.~Schillaci}
\author[f]{B.~Schmitt}
\author[m,u]{B.~Singari}
\author[a,i]{A.~Soliman}
\author[f]{T.~St.~Germaine}
\author[a]{B.~Steinbach}
\author[b]{R.~Sudiwala}
\author[h,c]{K.~L.~Thompson}
\author[b]{C.~Tucker}
\author[i]{A.~D.~Turner}
\author[w]{C.~Verg\`{e}s}
\author[j,v]{A.~G.~Vieregg}
\author[a]{A.~Wandui}
\author[i]{A.~C.~Weber}
\author[m]{J.~Willmert}
\author[a,c,d]{W.~L.~K.~Wu}
\author[h]{H.~Yang}
\author[j,l]{C.~Yu}
\author[f]{L.~Zeng}
\author[c]{C.~Zhang}
\author[a]{S.~Zhang}

\affil[a]{Department of Physics, California Institute of Technology, 1200 E. California Boulevard, Pasadena, CA 91125, USA}
\affil[b]{School of Physics and Astronomy, Cardiff University, Cardiff, CF24 3AA, United Kingdom}
\affil[c]{Kavli Institute for Particle Astrophysics and Cosmology, Stanford University, Stanford, CA 94305, USA}
\affil[d]{SLAC National Accelerator Laboratory, Menlo Park, CA 94025, USA}
\affil[e]{Department of Physics and Astronomy, University of British Columbia, Vancouver, British Columbia, V6T 1Z1, Canada}
\affil[f]{Center for Astrophysics $|$ Harvard \& Smithsonian, Cambridge, MA 02138, USA}
\affil[g]{Department of Physics, University of Cincinnati, Cincinnati, OH 45221, USA}
\affil[h]{Department of Physics, Stanford University, Stanford, California 94305, USA}
\affil[i]{Jet Propulsion Laboratory, California Institute of Technology, 4800 Oak Grove Drive, Pasadena, CA 91109, USA}
\affil[j]{Kavli Institute for Cosmological Physics, University of Chicago, 5640 S Ellis Ave, Chicago, IL 60637, USA}
\affil[k]{National Institute of Standards and Technology, Boulder, CO 80305, USA}
\affil[l]{High-Energy Physics Division, Argonne National Laboratory, 9700 South Cass Avenue., Lemont, IL, 60439, USA}
\affil[m]{School of Physics and Astronomy, University of Minnesota, Minneapolis, MN 55455, USA}
\affil[n]{Service des Basses Temperatures, Commissariat a l’Energie Atomique, 38054 Grenoble, France}
\affil[o]{School of Engineering and Natural Sciences, University of Iceland, Sæmundargata 2, 102 Reykjavík, Iceland}
\affil[p]{Department of Physics, Harvard University, Cambridge, MA 02138, USA}
\affil[q]{Department of Physics, University of Illinois at Urbana-Champaign, Urbana, Illinois 61801, USA}
\affil[r]{Centre for Theoretical Cosmology, DAMTP, University of Cambridge, Wilberforce Road, Cambridge CB3 0WA, UK}
\affil[s]{Kavli Institute for Cosmology Cambridge, Madingley Road, Cambridge CB3 0HA, UK}
\affil[t]{Department of Physics, Boston University, 590 Commonwealth Avenue, Boston, MA 02215, USA}
\affil[u]{Minnesota Institute for Astrophysics, University of Minnesota, Minneapolis, MN 55455, USA}
\affil[v]{Department of Physics, University of Chicago, 5720 S Ellis Ave, Chicago, IL 60637, USA}
\affil[w]{Lawrence Berkeley National Laboratory, 1 Cyclotron Road, Berkeley, CA 94720, USA}

\authorinfo{Further author information: (Send correspondence to A.S.)\\A.S.: asteiger@caltech.edu}

\begin{document} 
\maketitle

\begin{abstract}
Measurements of the polarization of the cosmic microwave background are critical for modern cosmology as they constrain the physics of the early universe and cosmic inflation. BICEP Array is using a series of small-aperture polarimeters located at the South Pole to measure this signal with a projected $\sigma(r)\approx0.001$ by 2034. The 30/40 GHz receiver and the 150 GHz receiver have been observing the cosmic microwave background for multiple years, and data from these receivers will be included in the next published BICEP analysis result which includes all data taken through 2024. The 220/270 GHz receiver is partially completed with seven out of twelve detector modules installed, and is planned to be filled with five additional modules prior to the 2027 observation season. Finishing this receiver is a top priority for BA as the science goals of the experiment require the dust foreground cleaning this receiver will enable. On-site characterization efforts for this receiver from the 2025–26 austral summer show satisfactory spectroscopy, beam pointing, and 220 GHz efficiency, while the 270 GHz efficiency is an active area of work.
\end{abstract}


\section{INTRODUCTION}
\label{sec:intro}  % \label{} allows reference to this section
High-precision measurements of the cosmic microwave background (CMB) are crucial in understanding the physics of the early universe. The standard LCDM cosmological model predicts scalar perturbations (density waves) in the early universe that would give rise to E-modes in the polarization pattern of the CMB. Theories of cosmic inflation predict tensor perturbations (primordial gravitational waves) as a result of the rapid expansion of inflation, which in turn would cause B-modes in the CMB polarization as well. The tensor-to-scalar ratio $r$ constrains the energy scale of inflation, and so measuring these B-modes and thus measuring $r$ is of critical importance in modern cosmology\cite{CMBS4,Seljak_1997,Kamionkowski_1997,Seljak_1997_2}.

BICEP Array (BA) is the latest in the BICEP series of experiments, which have been measuring the polarization of the CMB with small-aperture polarimeters located at the South Pole since 2006. The small aperture design allows for high sensitivity at the degree scale on sky, which is the scale where the B-mode signal is expected to peak\cite{CMBS4}. The latest published result from the BICEP program consists of all BICEP data from 2010 through 2018 as well as external data from WMAP and Planck (BK18), and has $\sigma(r)=0.009$, which is the tightest constraint on $r$ to date\cite{BK18}. BA is a collection of BICEP3-style receivers\cite{Hui_2018,B3_inst,Nakato_2024}, henceforth referred to as BA-class receivers. This class of receiver is a significant upgrade over the \textit{Keck}-class receivers that made up BICEP2 and \textit{Keck} Array\cite{BICEP2,KA}. In addition to upgrades to the optics and cryogenics, the BA-class receivers have a significantly larger focal plane that allows for an order of magnitude increase in detector count. The next-generation sensitivity provided by BA will enable a deeper search for cosmic inflation than ever before, with a $\sigma(r)$ target of $\sim0.001$ by 2034. 

One of the reasons a deep search for CMB B-modes is difficult is the presence of polarized foregrounds that contaminate the CMB polarization signal. At low frequencies, synchrotron radiation is expected to be the dominant foreground and at high frequencies there is polarized galactic dust\cite{Foregrounds}. To combat the foreground problem, BA uses multi-frequency measurements well above and below the CMB peak of $\sim160$ GHz to characterize the foregrounds and clean them from the CMB signal.  The low frequency receiver designed to mitigate the synchrotron foreground has been observing since 2020, and the high frequency receiver designed to mitigate the dust foreground is nearing completion. Sample variance from dust accounts for more than 20\% of the uncertainty in the BK18 $r$ constraint\cite{BK18}, and so finishing this receiver is essential to the science goals of BA. 
\section{STATUS OF BICEP ARRAY AND BK24}
\label{BA_2026}

BA currently has two completed receivers, BA1-30/40 which observes at 30/40 GHz and BA2-150 which observes at 150 GHz, and a partially completed receiver called BA3-220/270 which observes at 220/270 GHz. BA1-30/40 was deployed in 2019; it houses 542 transition edge sensors (TES) and is most useful for constraining the synchrotron foreground. BA2-150 was deployed in 2022; it houses 7,776 TESs which provide high sensitivity in an optimal CMB-observing band. BA3-220/270 was deployed in 2024; it currently houses 4,536 TESs and will house 7,776 when completed, which will be accomplished prior to the 2027 observing season.\footnote{BA1-30/40 has fewer detectors than the other two receivers as the antenna size at this frequency is much bigger.} This receiver will be crucial in constraining the polarized galactic dust foreground\cite{Foregrounds}.

Data from the two fully completed receivers are part of the BK24 dataset, which includes all data from the BICEP/\textit{Keck} series of experiments through 2024. The BK24 data products shown in the remainder of this section are preliminary; a complete BK24 manuscript is currently in preparation.
\subsection{CMB Maps}
\label{CMB Maps}

This section presents CMB maps in the Stokes parameters T, Q, and U for the two BA receivers in the BK24 dataset.  The BA-class receivers have a larger instantaneous field of view than the \textit{Keck}-class receivers, which allows for more mode coverage per map and increased overall foreground coverage. In this section and the following sections, `large field' indicates a BA-class receiver, and `small field' indicates a \textit{Keck}-class receiver.

The first set of maps is Fig. \ref{map40}, which come from five years of BA1-30/40 40 GHz data. The additional map images can be found in appendix \ref{Map_Images}.
   \begin{figure} [h!]
   \begin{center}
   \begin{tabular}{c} %% tabular useful for creating an array of images 
   \includegraphics[height=10cm]{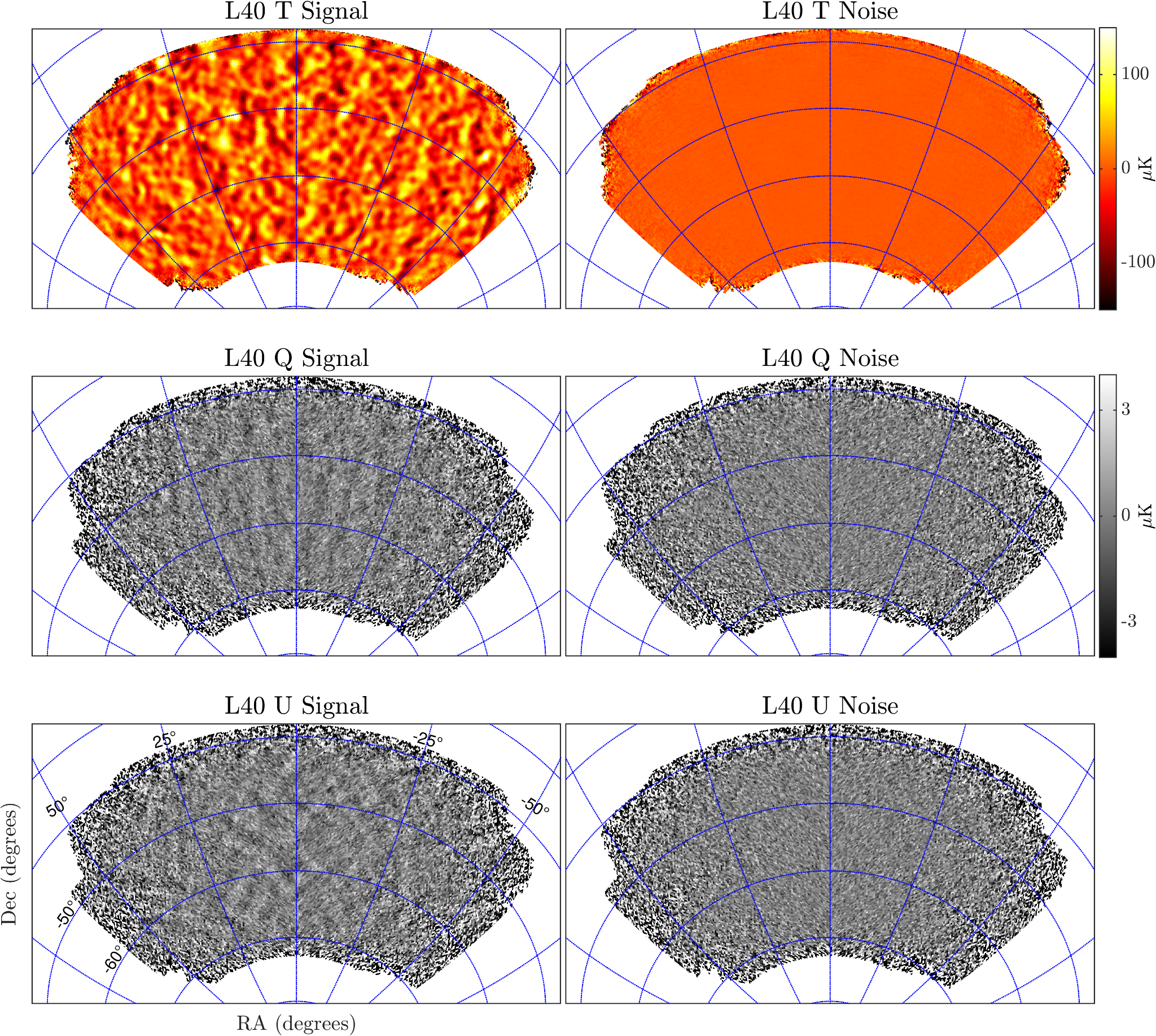}
   \end{tabular}
   \end{center}
   \caption[] 
%>>>> use \label inside caption to get Fig. number with \ref{}
   { \label{map40} 
CMB maps at 40 GHz from BA1-30/40. `L40' refers to large field at 40 GHz.}
   \end{figure} 
Although the 40 GHz dataset's primary intended use is constraining the synchrotron foreground, you can see the E-mode dominated LCDM signal as a plus pattern in Q and a cross pattern in U. Figure~\ref{map150} shows the maps from two years of BA2-150 data. The CMB is much brighter at 150 GHz compared to 40 GHz, and so the plus pattern in Q and the cross pattern in U are both much clearer. Figure \ref{map100} shows the maps from nine years of BICEP3 data. These are the deepest polarization maps ever at this frequency and scale with a map depth on the order of 1 $\mu$K-arcmin. The signal maps look similar to Fig. \ref{map150}, but the noise maps make the difference apparent. 

Figures \ref{map220} and \ref{map270} show the 220 GHz and 270 GHz maps from the \textit{Keck} Array receivers. This is where the 220 GHz and 270 GHz data for BK24 will come from, as the maps from BA3-220/270 are still being analyzed. These maps show the CMB patterns as well, but there is some blurring evident in the bottom right of the Q and U signal maps for both frequencies. This is polarized galactic dust, and one of the primary reasons receivers are deployed at these higher frequencies is to constrain this foreground. The dust pattern visibly changes between 220 and 270 GHz, demonstrating the need for multi-frequency dust measurements to understand the spatial and spectral distributions. 

The `S' in the sub-figure titles indicates these are small field maps, which are $\sim200$ sq. degrees smaller than the large field maps. These small field maps allow for cleaning the dust from the other small field maps, but to clean the dust from the maps in Figs. \ref{map100} and \ref{map150}, large field dust maps are required. BA3-220/270 will produce these larger maps.
   
\subsection{BK24 Noise Levels}
\begin{figure} [h]
   \begin{center}
   \begin{tabular}{c} %% tabular useful for creating an array of images 
   \includegraphics[height=12cm]{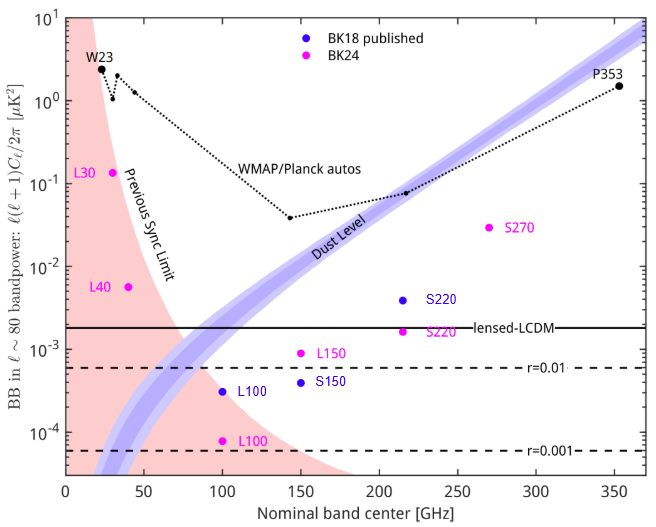}
   \end{tabular}
   \end{center}
   \caption[] 
%>>>> use \label inside caption to get Fig. number with \ref{}
   { \label{noise_plot} 
Noise uncertainties of the B-mode autospectra at $\ell\sim80$ for BK18 (blue) and BK24 (pink), and external maps (WMAP and Planck) used in BK18 (black). The horizontal continuous and dashed black lines show the sensitivity levels required to measure the lensed-LCDM signal, and the r=0.01 and r=0.001 B-mode signals. The synchrotron upper limit and dust foregrounds are shown as the pink shaded region and blue strip, respectively.}
   \end{figure}

Figure \ref{noise_plot} shows the preliminary BK24 B-mode autospectra noise uncertainties at $\ell\sim80$, which corresponds to $\sim2.25^{\circ}$ scale on sky. Also shown are the noise levels from BK18 from both BICEP data and external maps from WMAP and Planck that were included to help constrain foregrounds. The synchrotron upper limit and dust levels are included, as well as the sensitivity levels needed to measure the lensed-LCDM signal, and the r=0.01 and r=0.001 B-mode signals.

At the lower frequencies, the large field 30 and 40 GHz maps from BA1-30/40 are below the synchrotron limit, implying that the synchrotron foreground constraints in BK24 will be driven by BA data rather than external WMAP and Planck observations as in BK18. The BICEP3 `L100' data point is the deepest map in both the BK18 and BK24 dataset. At 150 GHz, the large field map from BA2-150 is quickly approaching the sensitivity level of the combined \textit{Keck} Array / BICEP2 data despite only two years of observing time, demonstrating the improvement in mapping speed obtained from upgrading from a \textit{Keck}-class receiver to a BA-class receiver. The 220 and 270 GHz data from the \textit{Keck} array have the highest sensitivity to dust in the BK24 dataset, and can be used to clean dust from the small field maps, but cleaning the large field maps and the science goals of the BICEP program require large field maps at these frequencies with higher sensitivities to dust than those of the `S220' and `S270' maps. Achieving these low noise levels at these higher frequencies is the purpose of BA3-220/270.

\subsection{Projections Beyond BK24}

\begin{figure} [h!]
   \begin{center}
   \begin{tabular}{c} %% tabular useful for creating an array of images 
   \includegraphics[height=12.7cm]{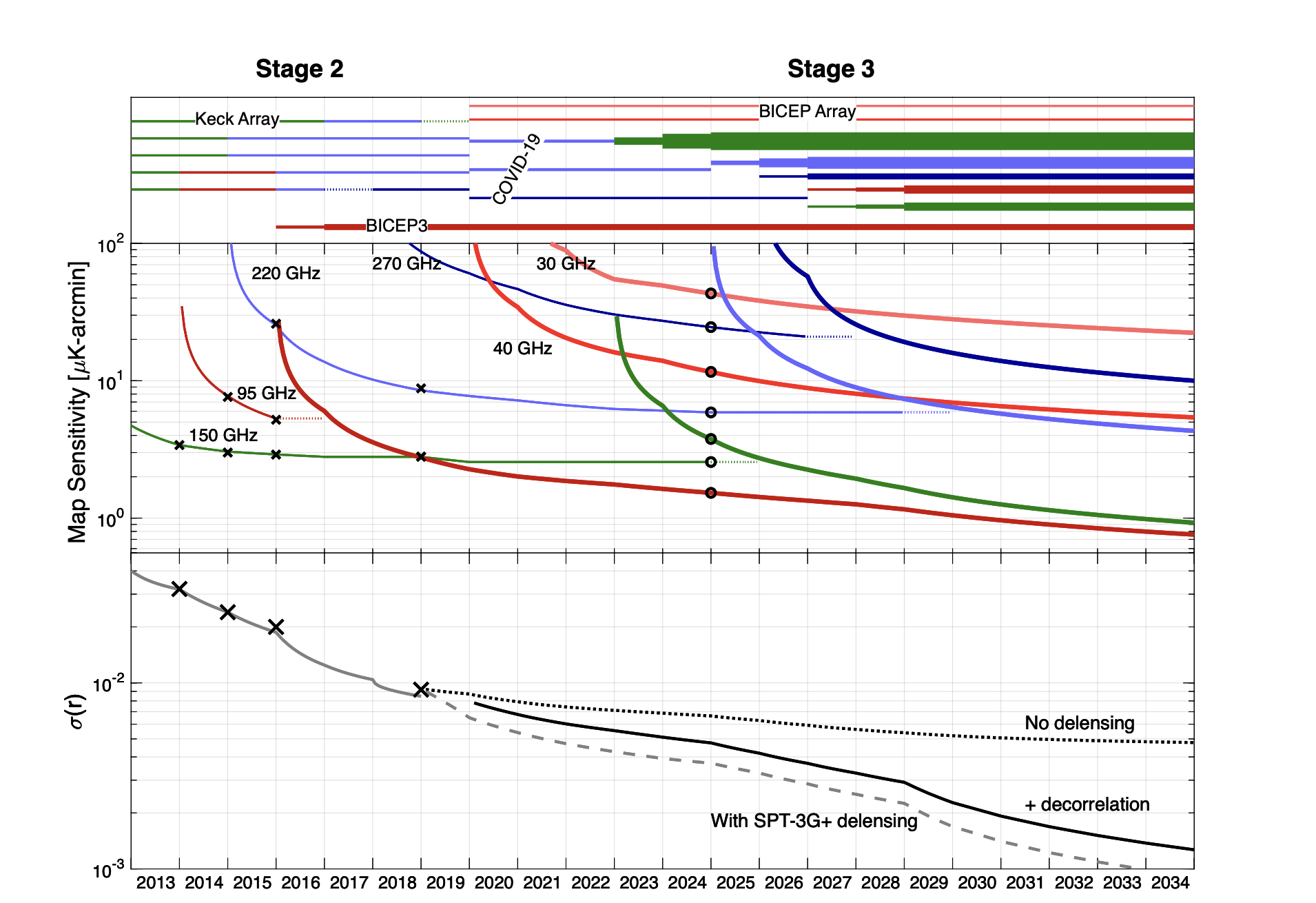}
   \end{tabular}
   \end{center}
   \caption[] 
%>>>> use \label inside caption to get Fig. number with \ref{}
   { \label{projections_plot} 
Top: BICEP experimental configuration from 2013 to 2034 (planned). Each line is a receiver-frequency combination. The color denotes the frequency as denoted in the middle panel. The thickness denotes detector count. 
Middle: Map sensitivity per frequency from 2013 to 2034. A thin line corresponds to a \textit{Keck} class-receiver and a thick one corresponds to a BA-class receiver, and color denotes frequency. The X's represent BICEP publications, and the circles are the preliminary BK24 values. Everything past 2024 is a projection. 
Bottom: $\sigma(r)$ from 2013 to 2034, again X's are BICEP publications. Three scenarios are included, one with no delensing (small dashes), one with delensing and a foreground model allowing for line of sight dust decorrelation (continuous), and one with delensing and no decorrelation (large dashes).}
   \end{figure}

Figure \ref{projections_plot} shows projections for the BICEP program beyond BK24. The top panel shows the experimental configuration from 2013 to 2034, with future years according to the program plan for finishing BICEP Array. Each line is a receiver/frequency combination (Ex. BA1-30/40 has two lines, one for 30 GHz and one for 40 GHz). The thickness of the line represents detector count, and the color represents frequency as denoted in the middle panel. A line getting thicker with each year corresponds to additional detectors being deployed as the receiver gets completed.

The middle panel shows the map sensitivity for each frequency as a function of time. A thin line corresponds to a \textit{Keck}-class receiver and a thick line corresponds to a BA-class receiver. The X's represent BICEP/\textit{Keck} publications and the circles are the preliminary BK24 values, and everything after 2024 is a projection. The 30 and 40 GHz sensitivities are sufficient for improving synchrotron constraints as shown in Fig. \ref{noise_plot}. At 95 GHz, BICEP3 replaced the \textit{Keck} receivers in 2016 and has been steadily observing the CMB since then. The BICEP3 preliminary BK24 map sensitivity is the best in the dataset at $\sim1$ $\mu{K}$-arcmin. At 150 GHz, the map sensitivity coming from just two years of BA2-150 observations is close to that of the 18 receiver years of \textit{Keck} Array and BICEP2 combined, and is projected to surpass that value in 1–2 years. The case with the higher frequencies is similar with the sensitives at 220 and 270 GHz from BA3-220/270 projected to surpass those of the \textit{Keck} receivers after about five years of observations.

The bottom panel shows $\sigma(r)$ as a function of time; again the X's indicated BICEP/\textit{Keck} publications. The value of $\sigma(r)$ post BK18 strongly depends on effective delensing. The BICEP program and the South Pole Telescope (SPT) experiment have partnered to form the South Pole Observatory, where high resolution SPT maps are being used to de-lens BICEP data\cite{Lensing}. An SPO forecasting paper that discusses these forecasts in greater detail is in preparation. Three projection scenarios are shown, one with no delensing, one with delensing and a foreground model that allows for dust decorrelation,\footnote{Dust that is not perfectly correlated across frequencies for a given line of sight.} and one with no delensing and no decorrelation. The projection scenario with effective delensing and no decorrelation projects that the experiment will achieve its goal of $\sigma(r)\sim0.001$ by 2034. Previous BICEP/\textit{Keck} projections have a strong track record of matching performance as they are based on achieved performance and conservative assumptions\cite{proj_proof}.

\section{STATUS OF BA3-220/270}
\label{BA3}
\begin{figure} [h!]
   \begin{center}
   \begin{tabular}{c} %% tabular useful for creating an array of images 
   \includegraphics[height=8cm]{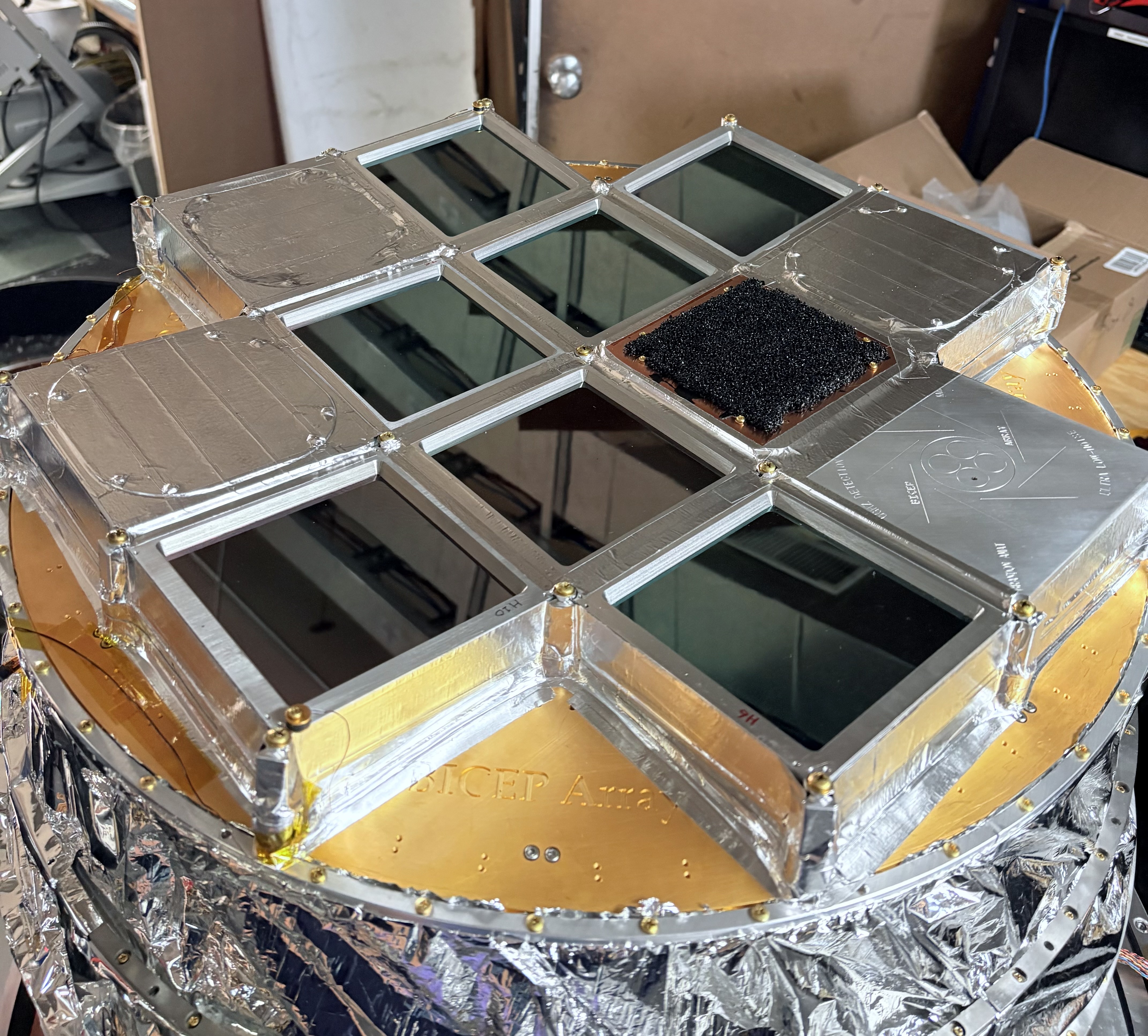}
   \end{tabular}
   \end{center}
   \caption[] 
%>>>> use \label inside caption to get Fig. number with \ref{}
   { \label{ba3_fp} 
BA3-220/270 focal plane with seven modules installed.}
   \end{figure}

When completed, BA3-220/270 will allow BA to constrain the polarized galactic dust foreground at the levels required to reach $\sigma(r)\sim0.001$. Currently, the receiver has seven out of twelve detector modules installed, six at 220 GHz and one at 270 GHz. Filling out the rest of the focal plane is a top priority for BA, and the nominal plan is to finish the focal plane prior to the 2027 observation season with two more 220 GHz modules and three more 270 GHz modules. An image of the BA3-220/270 with the seven installed modules is shown in Fig. \ref{ba3_fp}.
The following subsections describe the on-site measurements done during the 2025–26 austral summer to characterize the receiver.

\subsection{Optical Efficiency}
\begin{figure} 
   \begin{center}
   \begin{tabular}{c} %% tabular useful for creating an array of images 
   \includegraphics[height=8.8cm]{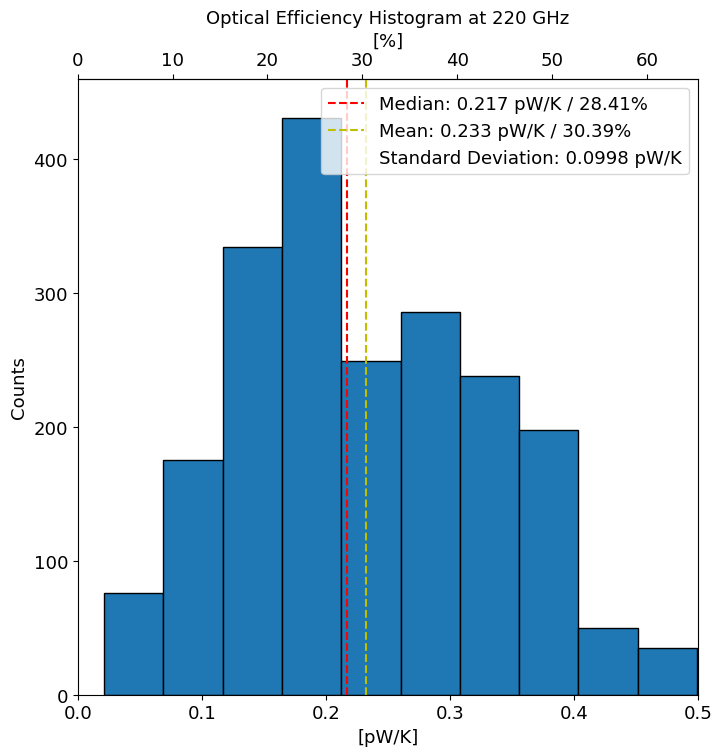}
   \includegraphics[height=8.8cm]{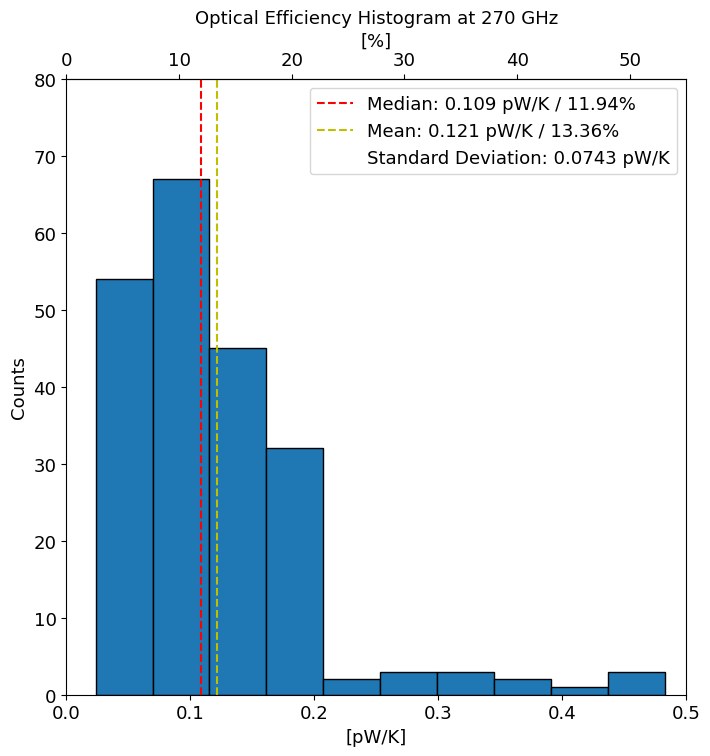}    
   \end{tabular}
   \end{center}
   \caption[] 
%>>>> use \label inside caption to get Fig. number with \ref{}
   { \label{oe_hists} 
Optical efficiency histograms for 220 GHz (left) and 270 GHz (right).}
   \end{figure}

Optical efficiency is calculated by comparing the response of the TESs to a 300K load (room temperature) and a 77K load (liquid nitrogen). For each thermal load, we produce a power vs. resistance curve for each of the detectors, and from this we calculate a saturation power. Then efficiency is calculated as:
$$
\frac{dP}{dT}=\frac{P_{sat,300K}-P_{sat,77K}}{300K-77K}
$$
The per-module median efficiencies are shown in Tab. \ref{oe_table} in both physical units and percent efficiency.\footnote{Percent efficiencies are calculated with a $\sim$55 GHz bandwidth at 220 GHz and $\sim$65 GHz bandwidth at 270 GHz. See Sec. \ref{sub_fts}.} Per-frequency efficiency histograms are shown in Fig. \ref{oe_hists}. 

\begin{table}[ht]
\caption{BA3-220/270 Optical Efficiency} 
\label{oe_table}
\begin{center}       
\begin{tabular}{|l|l|l|} 
\hline
\rule[-1ex]{0pt}{3.5ex}  Module/Frequency & Optical Efficiency (pW/K) & Optical Efficiency (\%)  \\
\hline
\rule[-1ex]{0pt}{3.5ex}  220 GHz \#1 & 0.17 & 22 \\
\hline
\rule[-1ex]{0pt}{3.5ex}  220 GHz \#2 & 0.20 & 26 \\
\hline
\rule[-1ex]{0pt}{3.5ex}  220 GHz \#3 & 0.29 & 38 \\
\hline
\rule[-1ex]{0pt}{3.5ex}  220 GHz \#4 & 0.16 & 21 \\
\hline 
\rule[-1ex]{0pt}{3.5ex}  220 GHz \#5 & 0.28 & 37 \\
\hline 
\rule[-1ex]{0pt}{3.5ex}  220 GHz \#6 & 0.36 & 47 \\
\hline 
\rule[-1ex]{0pt}{3.5ex}  270 GHz \#1 & 0.10 & 12 \\
\hline 
\end{tabular}
\end{center}
\end{table}

The 220 GHz population has an appreciable spread, but the optical efficiency is generally about $30\%$ with none of the modules having a median efficiency less than $20\%$. Noting that this test measures end-to-end efficiency and not isolated TES efficiency, these results are satisfactory. 

For the sole 270 GHz module in the receiver, the optical efficiency is about $12\%$. This is sub-optimal for BA's science goals, and this module will be replaced once all the empty module slots on the focal plane have been filled. In-lab testing prior to the deployment of this tile indicated the sub-optimal performance, but deploying and characterizing a 270 GHz tile on-site was beneficial. The 270 GHz detector tile design has received multiple changes since this module was fabricated, including a fix for a misalignment between the ground plane and device layer, which should increase efficiency. Devices produced using the improved fabrication procedure during the upcoming season will test whether this change resolves the issue.

\subsection{Fourier Transform Spectroscopy}
\label{sub_fts}
\begin{figure} 
   \begin{center}
   \begin{tabular}{c} %% tabular useful for creating an array of images 
   \includegraphics[height=8.5cm]{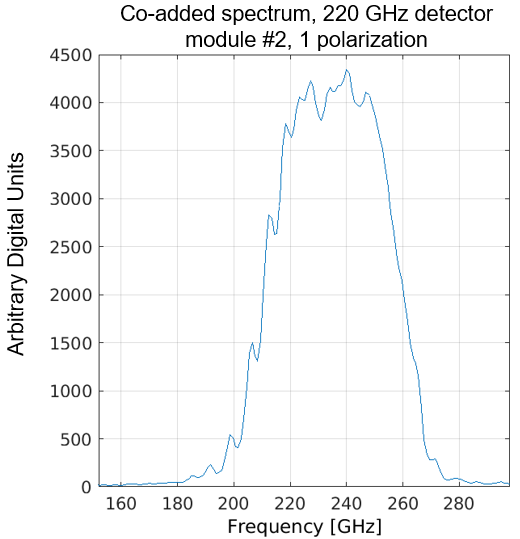}
   \includegraphics[height=8.5cm]{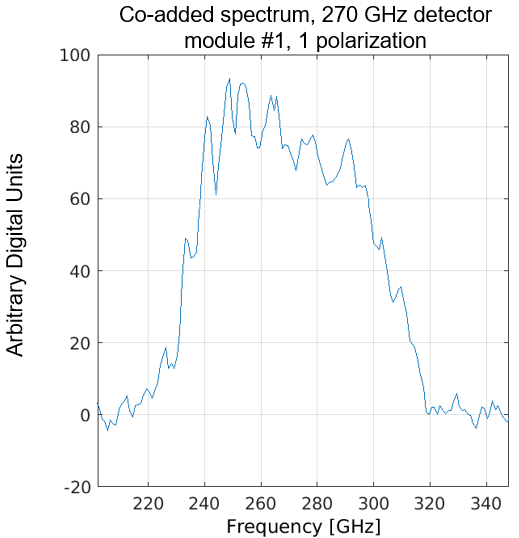}    
   \end{tabular}
   \end{center}
   \caption[] 
%>>>> use \label inside caption to get Fig. number with \ref{}
   { \label{fts_coadd} 
Co-added single-polarization spectra for 220 GHz module \#2 (left) and 270 GHz module \#1 (right).}
   \end{figure}

The spectra seen by the detectors are measured with a Martin–Pupplett Fourier Transform Spectrometer (FTS)\cite{MP_FTS}. Six total FTS measurements were taken of the focal plane with three pointings for each polarization to ensure full focal plane coverage. The per-module median bandcenters and bandwidths are recorded in Tab. \ref{fts_table}. Single polarization co-added spectra for one module/polarization from each frequency are shown in Fig. \ref{fts_coadd}.

\begin{table}[ht]
\caption{BA3-220/270 Module Median Bandcenters and Bandwidths} 
\label{fts_table}
\begin{center}       
\begin{tabular}{|l|l|l|} 
\hline
\rule[-1ex]{0pt}{3.5ex}  Module/Frequency & Bandcenter (GHz) & Bandwidth (GHz)  \\
\hline
\rule[-1ex]{0pt}{3.5ex}  220 GHz \#1 & 230.68 & 49.55 \\
\hline
\rule[-1ex]{0pt}{3.5ex}  220 GHz \#2 & 232.76 & 59.20 \\
\hline
\rule[-1ex]{0pt}{3.5ex}  220 GHz \#3 & 230.98 & 60.16 \\
\hline
\rule[-1ex]{0pt}{3.5ex}  220 GHz \#4 & 233.31 & 55.59 \\
\hline 
\rule[-1ex]{0pt}{3.5ex}  220 GHz \#5 & 234.31 & 59.24 \\
\hline 
\rule[-1ex]{0pt}{3.5ex}  220 GHz \#6 & 234.57 & 57.57 \\
\hline 
\rule[-1ex]{0pt}{3.5ex}  270 GHz \#1 & 266.69 & 65.57 \\
\hline 
\end{tabular}
\end{center}
\end{table}

The 220 GHz median bandwidth is 233 GHz. This is intentional and comes from a noise optimization of the band\cite{Filters}. The band is still referred to as 220 GHz for historical reasons. The 220 GHz median bandwidth is 57 GHz. The 270 GHz median bandcenter is 267 GHz and the median bandwidth is 66 GHz, and so both bands have a fractional bandwidth of $\sim0.245$. The 270 GHz FTS measurements yielded fewer detectors than the 220 GHz measurements, resulting in the co-added spectrum in Fig. \ref{fts_coadd} looking more featured. With higher quality 270 GHz modules, the co-added spectrum should look smoother. Overall, the FTS results from both bands represent no concerns.

\subsection{Far-Field Beam Mapping}
\begin{figure} 
   \begin{center}
   \begin{tabular}{c} %% tabular useful for creating an array of images 
   \includegraphics[height=8cm]{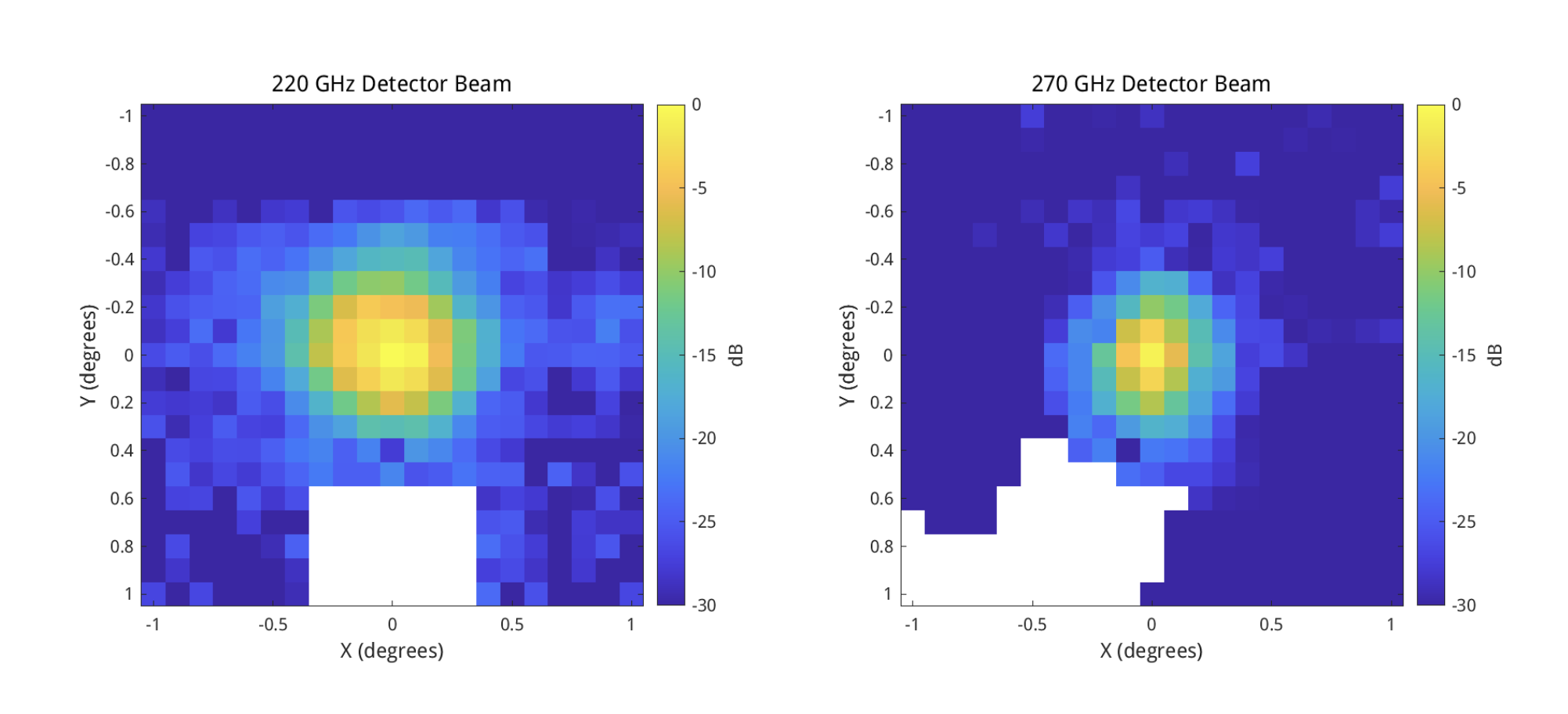}
   \end{tabular}
   \end{center}
   \caption[] 
%>>>> use \label inside caption to get Fig. number with \ref{}
   { \label{Beam_figs} 
Single far-field beam for 220 GHz detector (left) and 270 GHz detector (right).}
   \end{figure}

An extensive far-field beam mapping campaign using a flat mirror was carried out in early 2026. A full analysis of the data  similar to Ref.~\citenum{beams} is still in progress, but Fig.~\ref{Beam_figs} shows example beams for a detector from both frequencies.

The beams shown in Fig.~\ref{Beam_figs} come from a single measurement. Once the analysis is complete each detector will have multiple measurements and so the missing sections of the beams will be filled in and the beam maps will get substantially deeper. Regardless, the beams show the expected 2D-Gaussian shape with no glaring features, and they follow the expected $\lambda/D$ scaling. 
\section{CONCLUSION}
BA is utilizing a series of small-aperture polarimeters located at the South Pole to measure the polarization of the CMB with groundbreaking precision. Current projections show BA reaching its sensitivity goal of $\sigma(r)\approx0.001$ by 2034. BA1-30/40 and BA2-150 have been observing for multiple years now, and data from these receivers are being included in the BK24 dataset. In particular, BA2-150 is already approaching the depth of all previous 150 GHz data with just two years of observing, demonstrating the potential of the BA-class receiver. 

BA3-220/270 is partially completed and is expected to be fully completed prior to the 2027 observing season. Large field dust maps with the sensitivity provided by a BA-class receiver are critical to BA, and so finishing the focal plane of this receiver is a top priority. Regarding the seven modules installed in the receiver already, spectroscopy and beam mapping measurements present no concerns for both 220 GHz and 270 GHz. The 220 GHz efficiency is also acceptable, while the 270 GHz efficiency is an active area of work for BICEP and should be improved with the coming modules.
\newpage

\appendix
\renewcommand{\thefigure}{\Alph{section}-\arabic{figure}}
\renewcommand{\theHfigure}{\Alph{section}.\arabic{figure}}
\setcounter{figure}{0}
\section{ADDITIONAL CMB MAP IMAGES}
\label{Map_Images}

   \begin{figure} [h]
   \begin{center}
   \begin{tabular}{c} %% tabular useful for creating an array of images 
   \includegraphics[height=10cm]{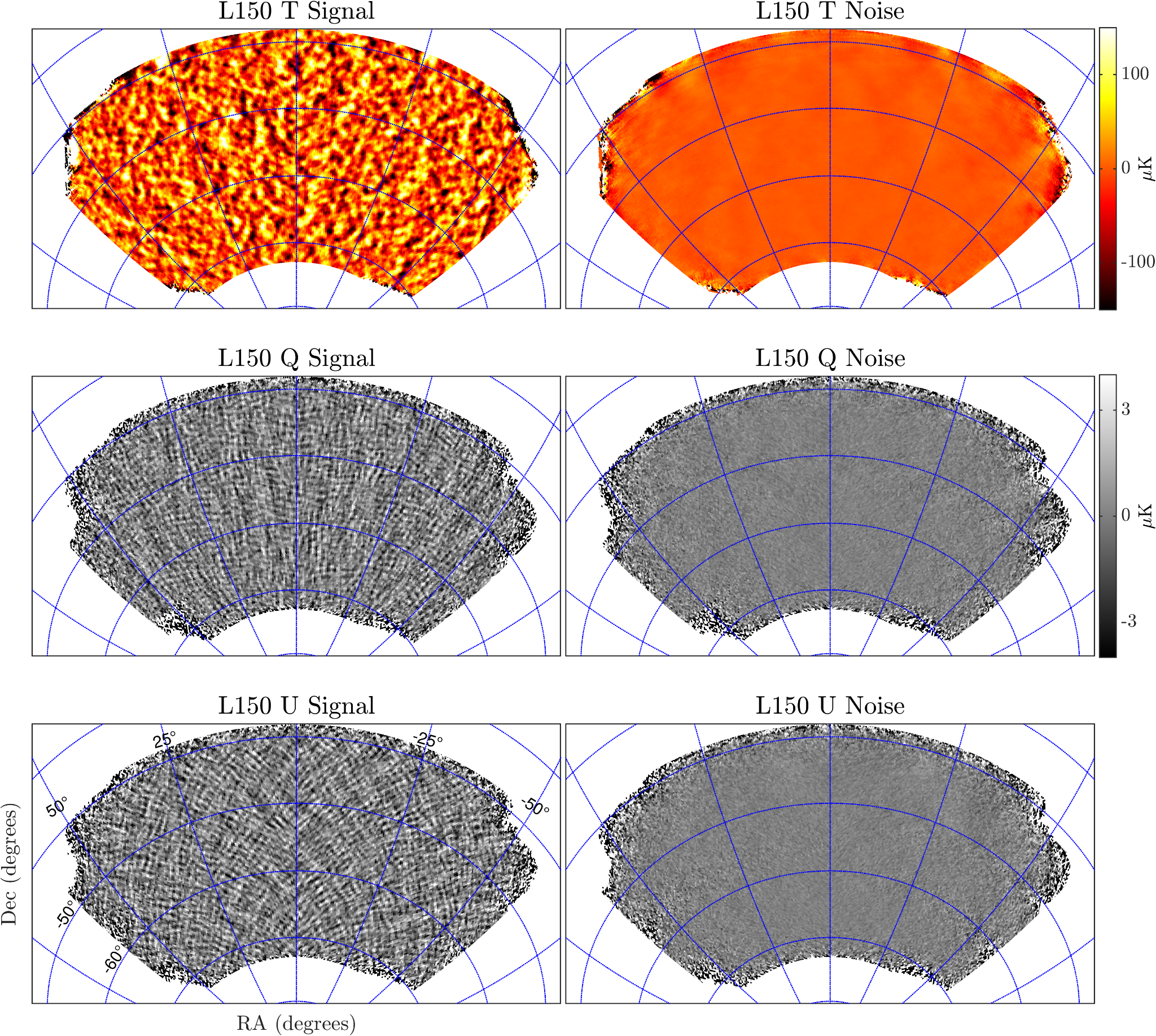}
   \end{tabular}
   \end{center}
   \caption[] 
%>>>> use \label inside caption to get Fig. number with \ref{}
   { \label{map150} 
   CMB maps from BA2-150. `L150' refers to large field at 150 GHz.}
   \end{figure} 

   \begin{figure} 
   \begin{center}
   \begin{tabular}{c} %% tabular useful for creating an array of images 
   \includegraphics[height=10cm]{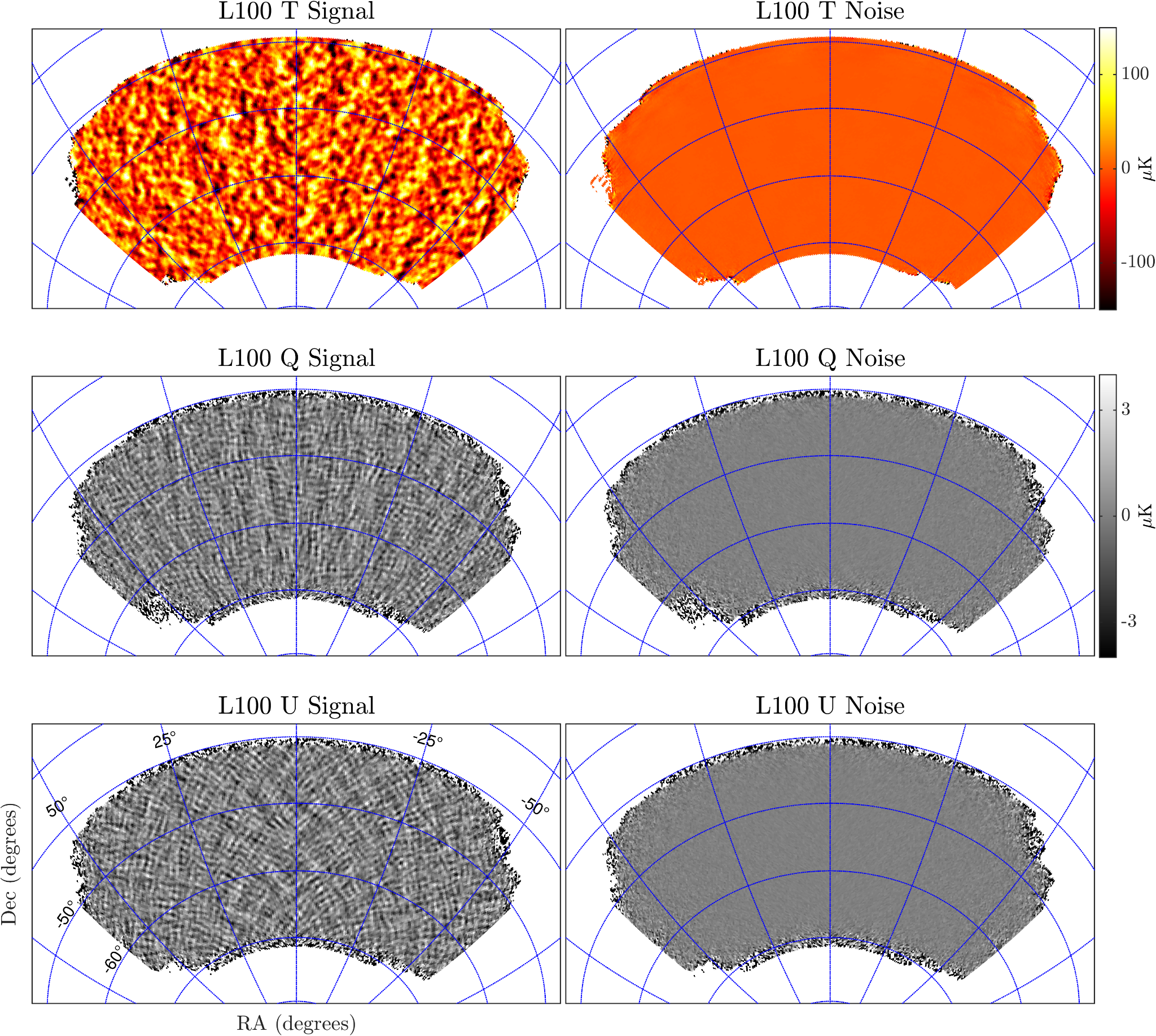}
   \end{tabular}
   \end{center}
   \caption[] 
%>>>> use \label inside caption to get Fig. number with \ref{}
   { \label{map100} 
CMB maps from BICEP3. `L100' refers to large field at 100 GHz.}
   \end{figure} 

   \begin{figure}
   \begin{center}
   \begin{tabular}{c} %% tabular useful for creating an array of images 
   \includegraphics[height=10cm]{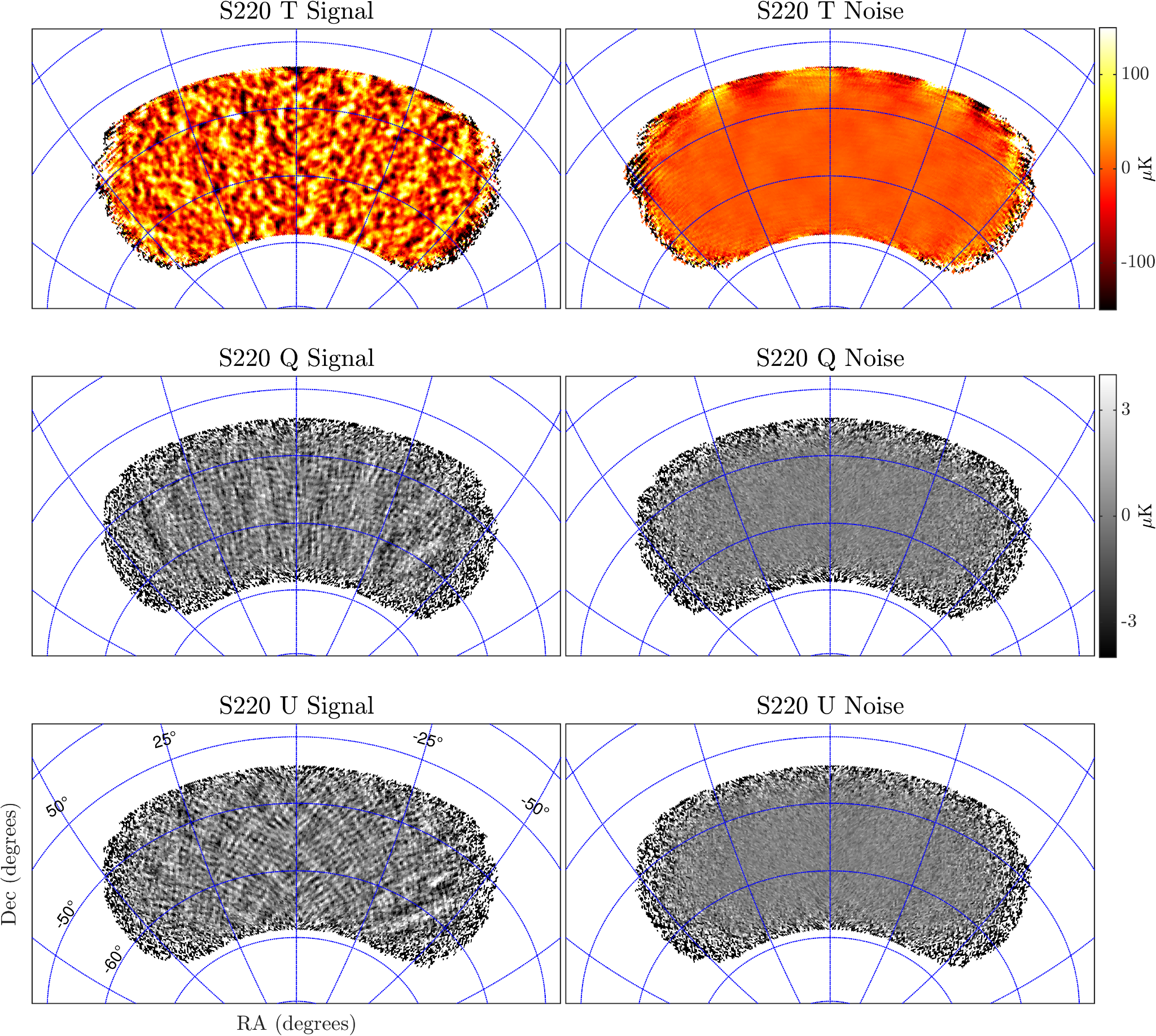}
   \end{tabular}
   \end{center}
   \caption[] 
%>>>> use \label inside caption to get Fig. number with \ref{}
   { \label{map220} 
CMB maps from the 220 GHz \textit{Keck} receivers. `S220' refers to small field at 220 GHz.}
   \end{figure}

   \begin{figure}
   \begin{center}
   \begin{tabular}{c} %% tabular useful for creating an array of images 
   \includegraphics[height=10cm]{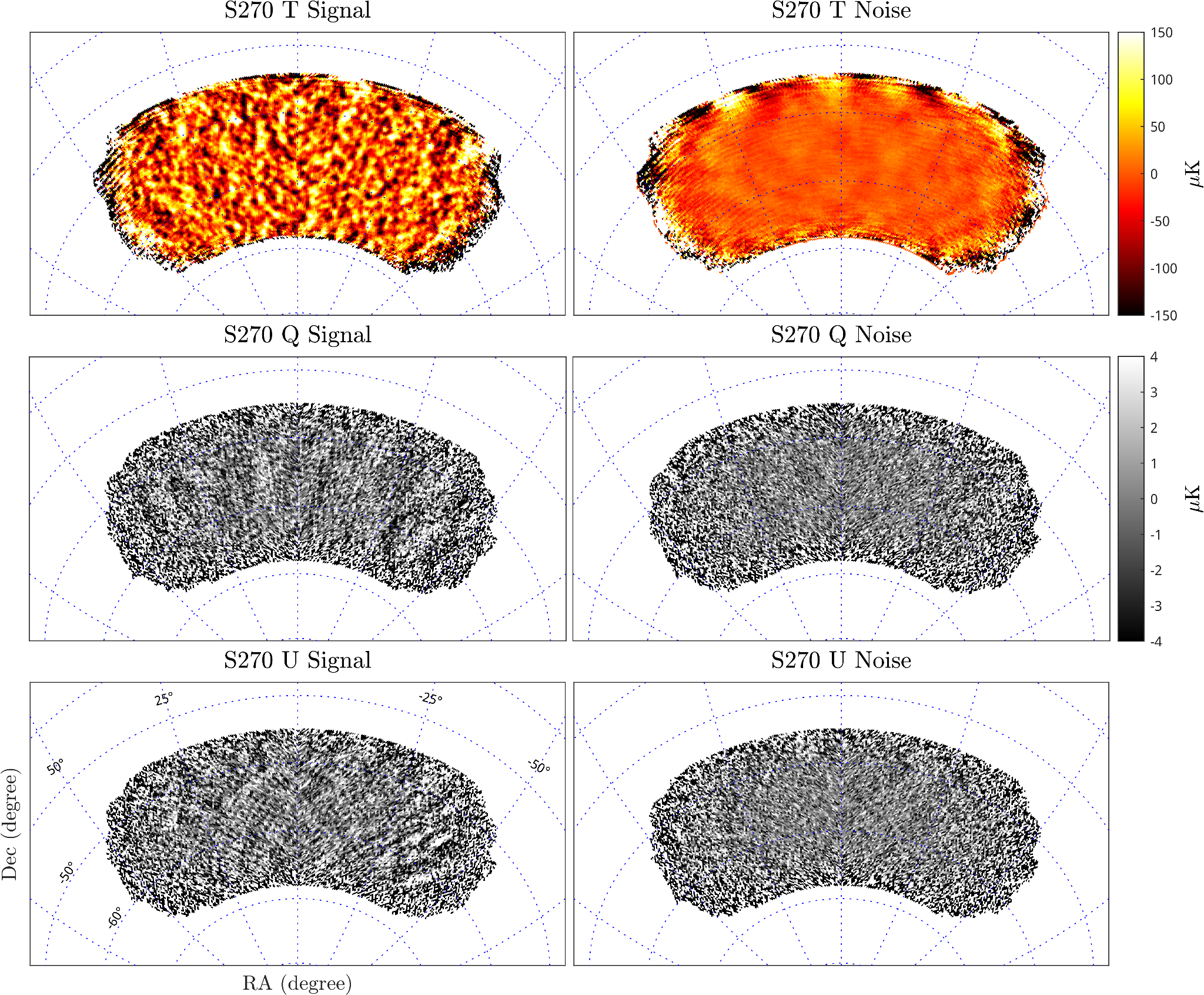}
   \end{tabular}
   \end{center}
   \caption[] 
%>>>> use \label inside caption to get Fig. number with \ref{}
   { \label{map270} 
CMB maps from the 270 GHz \textit{Keck} receivers. `S270' refers to small field at 270 GHz.}
   \end{figure} 
% References
\clearpage
\acknowledgments % equivalent to \section*{ACKNOWLEDGMENTS}       
 
The BICEP/\textit{Keck} experiments have been funded through
U.S. National Science Foundation grants most recently including 2220444-2220448, 2216223, 1836010, and 1726917.
The research was carried out at the Jet Propulsion Laboratory, California Institute of Technology, under a contract with the National Aeronautics and Space Administration (80NM0018D0004). Focal plane development and
testing were supported by the Gordon and Betty Moore
Foundation at the California Institute of Technology. Readout
electronics were supported by the Canada Foundation for
Innovation grant to the University of British Columbia. The
computations in this paper were run on the Cannon cluster
supported by the FAS Science Division Research Computing
Group at Harvard University. The analysis effort at Stanford
University and the SLAC National Accelerator Laboratory was
partially supported by the Department of Energy. We thank
the staff of the U.S. Antarctic Program and in particular the
South Pole Station without whose help this research would not
have been possible. We also thank our winter-over operators:
Manwei Chan, Karsten Look, Calvin Tsai, Paula Crock, Ta
Lee Shue, Grantland Hall, Hans Boenish, Robert Schwarz,
Sam Harrison, Anthony DeCicco, Thomas Leps, Brandon
Amat, Nathan Precup, Steffen Richter, Thibault Romand, Danielle Simmons, Markus Ayasse, Steven Jungst, Nathan McReynolds, and John Della Costa.

\bibliography{steiger_bib} % bibliography data in report.bib
\bibliographystyle{spiebib} % makes bibtex use spiebib.bst

\end{document}